\documentclass[a4paper,12pt]{article}

\usepackage[spanish]{babel} 
\usepackage{natbib}     
\usepackage[utf8]{inputenc} 
\usepackage{graphicx} 
\usepackage{amsmath} 
\usepackage{amssymb} 
\usepackage{cite} 
\usepackage{booktabs} 
\usepackage{hyperref} 
\usepackage{fancyhdr} 
\usepackage[left=2.5cm,right=2.5cm,top=2.5cm,bottom=2.5cm]{geometry} 

\hypersetup{
    colorlinks=true,
    linkcolor=blue,
    filecolor=magenta,      
    urlcolor=cyan,
    pdftitle={Enfoque Odychess},
    pdfauthor={Ernesto Giralt Hernandez, Lazaro Antonio Bueno Perez},
}

\title{\textbf{Enfoque Odychess: Un método dialéctico, constructivista y adaptativo para la enseñanza del ajedrez con inteligencias artificiales generativas}}
\author{Ernesto Giralt Hernandez\textsuperscript{1*} \\ \textit{https://orcid.org/0009-0005-6357-7522} \\
\\
\textsuperscript{1}AquarelleAI. Barcelona, España.\\
\\
Lazaro Antonio Bueno Perez\textsuperscript{2*} \\ \textit{https://orcid.org/0000-0002-5187-0968} \\
\\
\textsuperscript{2}Universidad de Camagüey, Facultad de Cultura Física. Camagüey, Cuba.\\
\\
\textsuperscript{*}Autor para la correspondencia: ernesto.giralt@aquarelleai.com}

\begin{document}

\maketitle

\section*{RESUMEN}
\textbf{Introducción:} La didáctica del ajedrez ha evolucionado a través de diferentes enfoques, sin embargo, las metodologías tradicionales, a menudo basadas en la memorización, contrastan significativamente con las nuevas posibilidades que ofrece la inteligencia artificial generativa, una tecnología aún poco explorada en este campo.

\textbf{Objetivo:} validar empíricamente la efectividad del Enfoque Odychess en la mejora del conocimiento ajedrecístico, la comprensión estratégica y las habilidades metacognitivas en estudiantes.

\textbf{Materiales y métodos:} Se llevó a cabo un estudio cuasi-experimental con un diseño pre-test/post-test y un grupo control (N=60). La intervención experimental implementó el Enfoque Odychess, incorporando un modelo de lenguaje Llama 3.3 que fue adaptado específicamente mediante técnicas de Parameter-Efficient Fine-Tuning (PEFT) para actuar como tutor ajedrecístico socrático. Se utilizaron instrumentos de evaluación cuantitativa para medir el conocimiento ajedrecístico, la comprensión estratégica y las habilidades metacognitivas antes y después de la intervención.

\textbf{Resultados:} Los resultados del estudio cuasi-experimental mostraron mejoras significativas en el grupo experimental en comparación con el grupo control en las tres variables analizadas: conocimiento ajedrecístico, comprensión estratégica y habilidades metacognitivas. El análisis cualitativo complementario reveló una mayor profundidad analítica, un razonamiento dialéctico más desarrollado y una mayor motivación intrínseca en los estudiantes que participaron en la intervención basada en el método Odychess.

\textbf{Conclusiones:} el Enfoque Odychess representa una metodología pedagógica eficaz para la enseñanza del ajedrez, demostrando el potencial de la integración sinérgica de principios constructivistas y dialécticos con la inteligencia artificial generativa. Las implicaciones de este trabajo son relevantes para educadores e instituciones interesadas en la adopción de tecnologías pedagógicas innovadoras y para investigadores en el campo de la IA aplicada a la educación, destacando la transferibilidad de la metodología de adaptación de modelos de lenguaje a otros dominios educativos.

\textbf{Palabras clave:} ajedrez educativo, inteligencia artificial en educación, aprendizaje dialéctico, tutoría adaptativa.

\section{INTRODUCCIÓN}
Diversos autores han abordado la enseñanza del ajedrez, como (Bueno, L. 2000; Ramírez, Bueno \& Gordo, 2016), fundamentándose principalmente en el constructivismo pedagógico. Esta teoría postula la necesidad de proporcionar a los estudiantes herramientas y apoyos temporales (andamiaje) para que construyan sus propios procedimientos en la resolución de problemas, modificando y ajustando sus esquemas de conocimiento a partir de la experiencia. En esencia, el constructivismo enfatiza el aprendizaje activo y reflexivo: el alumno integra nuevos saberes relacionándolos con sus conocimientos previos y reformula sus ideas al enfrentarse a desafíos. Este enfoque contrasta con modelos tradicionales centrados en la transmisión pasiva de contenidos por parte del profesor, donde el estudiante asume un rol más activo en su propio aprendizaje.

De esta manera, el estudiante construye su comprensión mediante la resolución de ejercicios, la experimentación con ideas y la discusión de sus razonamientos, lo cual se alinea con el principio constructivista de la ``construcción mutua del conocimiento'' entre educador y aprendiz (Coll et al., 2006).

No obstante, la didáctica del ajedrez a menudo carece de estrategias que favorezcan un proceso de enseñanza-aprendizaje óptimo (Reyes-Joa et al., 2020). En este sentido, resulta crucial explotar plenamente la dialéctica a través de conversaciones reflexivas sobre posiciones ajedrecísticas. Esto induce al alumno a justificar sus movimientos, anticipar refutaciones y comparar líneas alternativas, entrenando no solo conocimientos específicos del ajedrez, sino también la flexibilidad cognitiva y la apertura al cambio de opinión ante la evidencia. Estas competencias son propias del pensamiento dialéctico y esenciales para un jugador de ajedrez competente. En suma, la dimensión dialéctica asegura que el aprendizaje del ajedrez trascienda la memorización mecánica, convirtiéndose en un proceso de indagación crítica donde cada concepto se comprende en profundidad al contrastarlo con alternativas y resolver las discrepancias que surgen durante el razonamiento.

En este contexto, el empleo de enfoques contemporáneos como el Aprendizaje Basado en Problemas (ABP) y la adquisición de patrones en el dominio ajedrecístico se tornan relevantes. En la pedagogía general, el ABP (Problem-Based Learning, PBL) se define como una estrategia didáctica centrada en la presentación de problemas complejos y reales como motor para el aprendizaje de conceptos clave, en contraposición a la exposición directa de contenidos teóricos (Barrows, 1986; Savery, 2006).

En el panorama educativo actual, la búsqueda de metodologías innovadoras que promuevan un aprendizaje significativo y profundo constituye una prioridad constante (USFA, 2024). Sin embargo, la aplicación efectiva de estos beneficios al ámbito del ajedrez requiere enfoques pedagógicos que superen la mera instrucción memorística.

En este escenario, el presente artículo introduce el Enfoque Odychess, un método pedagógico para la enseñanza del ajedrez que integra principios dialécticos y constructivistas con herramientas de inteligencia artificial generativa (IA).

Esta propuesta metodológica emerge de la necesidad de actualizar la didáctica ajedrecística, tradicionalmente enfocada en la memorización de jugadas o la instrucción unidireccional, hacia un modelo más interactivo, personalizado y centrado en el estudiante. Dentro de Odychess, el proceso de aprendizaje se concibe como una travesía cognitiva de descubrimiento, donde el estudiante construye activamente su conocimiento ajedrecístico a través de la resolución de problemas, el diálogo reflexivo y la adaptación continua de la dificultad de las tareas.

Este enfoque se apoya en modelos de lenguaje generativos avanzados (como Gemini de Google, Claude de Anthropic o el modelo Llama 3.3 de Meta, utilizado en nuestra implementación) (Grattafiori et al., 2024) para proporcionar tutoría personalizada, guiando al estudiante de manera interactiva y ajustándose a su nivel de juego.

Desde una perspectiva didáctica, el método Odychess se justifica por la convergencia de varios objetivos: fomentar un aprendizaje significativo del ajedrez (más allá de la simple acumulación de aperturas o patrones tácticos), desarrollar el pensamiento crítico y dialéctico de los estudiantes a través del juego, y aprovechar el potencial de las inteligencias artificiales generativas para ofrecer una tutoría individualizada a gran escala.

Paralelamente, Odychess enfatiza la enseñanza basada en patrones, reconociendo que la habilidad en ajedrez se sustenta considerablemente en la memoria y el reconocimiento de patrones posicionales y tácticos. La ciencia cognitiva del ajedrez ha demostrado que los maestros de élite no necesariamente calculan más profundamente que jugadores menos experimentados, sino que reconocen configuraciones familiares que guían su pensamiento. En un estudio clásico, Chase y Simon (1973) encontraron que los jugadores expertos poseen un extenso catálogo de patrones almacenados en la memoria a largo plazo, lo que les permite recordar e identificar rápidamente configuraciones de piezas significativas, a diferencia de los novatos. De hecho, la maestría ajedrecística proviene de la interiorización de representaciones cada vez más sofisticadas del tablero y sus patrones característicos (De Groot, 1965; Gobet \& Simon, 1996). Por ello, el entrenamiento efectivo del ajedrez suele implicar la exposición a miles de problemas tácticos y posiciones típicas, para que el jugador desarrolle una ``percepción'' casi instantánea para reconocer oportunidades (por ejemplo, un patrón de mate, una debilidad estructural, etc.).

El enfoque Odychess incorpora esta idea mediante la práctica sistemática de patrones en un contexto problemático. Cada problema o escenario presentado está diseñado para resaltar uno o varios patrones estratégicos/tácticos específicos. Por ejemplo, un Escenario de Aprendizaje Basado en Problemas Descontextualizado (EAPBD) podría centrarse en el patrón de clavada (pieza inmovilizada en una línea), otro en un típico mate de dama y caballo, y otro en un motivo estratégico como un peón aislado. Al resolver numerosos casos, el estudiante va formando en su memoria esquemas estructurados que posteriormente podrá evocar durante una partida real. Como señala el Gran Maestro Serper (2015), ``cuantos más patrones conoces, más fácil es encontrar buenas jugadas en tus partidas''.

La sinergia entre el ABP y la enseñanza basada en patrones en Odychess radica en que los problemas presentados proporcionan un contexto significativo para la adquisición de patrones. En lugar de estudiar una lista de temas tácticos de forma aislada, el alumno se enfrenta a ellos \textit{in situ} al resolver ejercicios. Esto conduce a un aprendizaje más profundo: el patrón deja de ser un concepto abstracto y se convierte en una herramienta concreta que le permitió superar un obstáculo. Además, la variedad de problemas, muchos de ellos descontextualizados de secuencias conocidas, evita que el estudiante dependa de la memorización rutinaria; lo fuerza a analizar la esencia de la posición y transferir principios conocidos a contextos nuevos.

Así, lo tratado por Bueno (2016) al proponer que el binomio problemas--patrones asegura un aprendizaje ajedrecístico significativo y transferible: los estudiantes adquieren repertorios de esquemas (patrones) y también la habilidad de aplicarlos flexiblemente resolviendo situaciones inéditas.

Los objetivos centrales del enfoque incluyen mejorar la comprensión estratégica y táctica del alumno, potenciar sus habilidades de resolución de problemas en contextos ajedrecísticos novedosos y desarrollar competencias metacognitivas (reflexión sobre su propio proceso de pensamiento) mediante la interacción con un tutor IA.

En última instancia, Odychess aspira a transformar la enseñanza del ajedrez en un proceso más dinámico, adaptativo y formativo, sirviendo de base para manuales de aplicación y guías para docentes que deseen incorporar esta metodología en sus prácticas. Es por ello que el objetivo de la presente investigación es validar empíricamente la efectividad del Enfoque Odychess en la mejora del conocimiento ajedrecístico, la comprensión estratégica y las habilidades metacognitivas en estudiantes.

\section{MATERIALES Y MÉTODOS}
Para evaluar la efectividad del Enfoque Odychess, se diseñó y ejecutó un estudio cuasi-experimental con un diseño pre-test/post-test y grupo de control no equivalente.

\begin{itemize}
    \item \textbf{Participantes:} La muestra estuvo compuesta por N=60 estudiantes de educación secundaria (edades 13-15 años) de dos centros educativos con características socioeconómicas similares. Los participantes tenían un nivel de ajedrez principiante-intermedio (ELO estimado $<$ 1200 FIDE) y se asignaron a las condiciones según los grupos-clase preexistentes (asignación no aleatoria de grupos intactos): Grupo Experimental (GE, n=30) que recibió instrucción mediante el Enfoque Odychess, y Grupo de Control (GC, n=30) que siguió un método de enseñanza de ajedrez tradicional basado en lecciones expositivas y resolución de ejercicios estándar. Se obtuvo consentimiento informado de los padres o tutores legales y el asentimiento de los estudiantes.

    \item \textbf{Intervención:} El GE participó en un programa de ajedrez basado en Odychess durante un semestre académico (16 semanas), con dos sesiones semanales de 60 minutos. Las sesiones fueron dirigidas por un docente capacitado en el enfoque Odychess y utilizaron la plataforma digital que integraba al tutor ``Odychess-Tutor''. Los estudiantes interactuaron con EAPBDs, participaron en diálogos socráticos con el docente y/o la IA, analizaron partidas y mantuvieron un portafolio digital. El GC recibió el mismo número de horas de instrucción, pero con un método tradicional centrado en la explicación de aperturas, medio juego y finales por parte del docente, seguido de la resolución de problemas tácticos y partidas de práctica sin el componente dialógico-adaptativo ni el uso del tutor IA específico.

    \item \textbf{Instrumentos de Recolección de Datos:}
    \begin{itemize}
        \item \textbf{Prueba de Conocimiento Ajedrecístico (PCA):} Un test compuesto por problemas tácticos estandarizados (evaluación de cálculo y reconocimiento de patrones), preguntas sobre principios estratégicos y evaluación de posiciones. Se administró al inicio (pre-test) y al final (post-test). La fiabilidad medida mediante alfa de Cronbach fue de 0.83.
        
        \item \textbf{Evaluación de Comprensión Estratégica (ECE):} Tarea basada en el análisis escrito de una posición compleja, evaluando la capacidad de formular planes, anticipar respuestas y justificar decisiones. Calificada mediante rúbrica por dos evaluadores independientes con un índice de concordancia inter-evaluadores kappa $>$ 0,85. Aplicada en pre-test y post-test.
        
        \item \textbf{Escala de Habilidades Metacognitivas en Ajedrez (EHMA):} Cuestionario tipo Likert adaptado para medir la autopercepción de planificación, monitoreo y evaluación del propio pensamiento durante el juego (basado en Schraw y Dennison, 1994). Aplicada en pre-test y post-test.
        
        \item \textbf{Registro de Interacciones con la IA:} Para el GE, se analizaron cualitativamente las transcripciones anonimizadas de los diálogos con el Odychess-Tutor, buscando evidencia de razonamiento dialéctico y reflexión.
        
        \item \textbf{Portafolios de Aprendizaje y Observación del Docente:} Análisis cualitativo de los portafolios (partidas comentadas, reflexiones) y notas de observación del docente sobre la participación y actitudes de los estudiantes en ambos grupos.
        
        \item \textbf{Cuestionario de Motivación y Compromiso (CMC):} Escala Likert midiendo interés, disfrute y percepción de utilidad de las clases de ajedrez. Aplicada al final del estudio.
    \end{itemize}

    \item \textbf{Procedimiento:} Tras la administración del pre-test (PCA, ECE, EHMA), se implementó la intervención durante 16 semanas. El docente del GE facilitó las sesiones Odychess, mientras otro docente impartió las clases tradicionales al GC. Se controló la equivalencia de experiencia y formación entre ambos docentes para minimizar este factor como variable extraña. Al finalizar el periodo, se administraron los post-tests (PCA, ECE, EHMA) y el CMC. Se recopilaron los datos de interacciones con la IA y los portafolios para análisis cualitativo.

    \item \textbf{Análisis de Datos:} Se utilizaron pruebas t de Student para muestras independientes para comparar las ganancias (post-test - pre-test) entre GE y GC en las medidas cuantitativas (PCA, ECE, EHMA). Se verificó previamente el cumplimiento de los supuestos de normalidad mediante la prueba de Shapiro-Wilk y la homogeneidad de varianzas con la prueba de Levene. Se aplicó análisis de covarianza (ANCOVA) para controlar posibles diferencias iniciales en el pre-test, utilizando las puntuaciones pre-test como covariables (Miller y Chapman, 2001). Para los datos cualitativos (interacciones con la IA, portafolios, observaciones), se realizó un análisis temático utilizando codificación abierta y axial para identificar patrones de pensamiento y actitudes (Braun y Clarke, 2006). Las respuestas al CMC se compararon mediante pruebas t. El nivel de significación estadística se estableció en p $<$ 0,05, y se calcularon los tamaños del efecto mediante la d de Cohen para determinar la magnitud de las diferencias.
\end{itemize}

\section{RESULTADOS Y DISCUSIÓN}

\subsection{Descripción del Ciclo Dialéctico-Adaptativo de enseñanza-aprendizaje}

El proceso de enseñanza-aprendizaje en Odychess se desarrolla mediante un Ciclo Dialéctico-Adaptativo iterativo, que asegura tanto la construcción dialógica del conocimiento como la adaptación continua al nivel del aprendiz (figura 1).

Este ciclo se repite a diferentes escalas (desde la micro-interacción en un solo problema hasta la planificación a lo largo de múltiples sesiones) y consta de varias fases interrelacionadas:

\begin{figure}[ht]
    \centering
    \includegraphics[width=\textwidth]{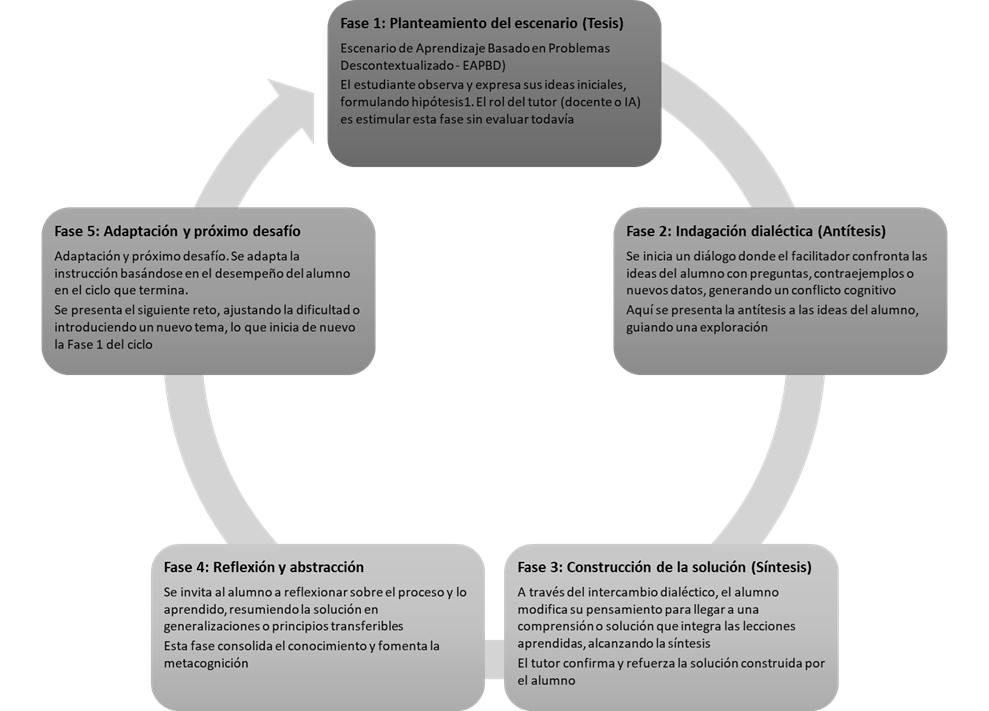}
    \caption{Ciclo dialéctico-adaptativo de enseñanza-aprendizaje en el Enfoque Odychess}
    \label{fig:ciclo}
\end{figure}

\begin{itemize}
    \item \textbf{Fase 1: Planteamiento del escenario (Tesis):} El ciclo inicia con la presentación de un desafío o problema (generalmente un EAPBD, como se describió). Aquí se expone la tesis, es decir, la situación inicial con la que el estudiante se confronta. Se brinda al alumno tiempo para observar la posición o enunciado y se le invita a expresar sus ideas iniciales: ¿Qué interpreta que está sucediendo? ¿Qué objetivo tendría que lograr? ¿Qué jugadas relevantes identifica? En esta fase el estudiante activa sus conocimientos previos pertinentes, formula hipótesis de solución o identifica lo que le resulta problemático. El rol del docente/IA es estimular esta fase preguntando: ``¿Qué elemento llama su atención?'', ``¿Cuál considera que es el problema a resolver?'', pero sin evaluar ni corregir todavía; se busca que el alumno comprometa una idea inicial (su tesis).

    \item \textbf{Fase 2: Indagación dialéctica (Antítesis):} Luego se desencadena el diálogo dialéctico propiamente dicho. El facilitador (humano o IA) confronta las ideas del alumno con preguntas, contraejemplos o nuevos datos, generando un conflicto cognitivo. Si el alumno planteó una solución equivocada o incompleta, aquí se le presenta la antítesis. Por ejemplo: ``Comprendo su plan de mate con la torre por la columna, pero ¿qué sucedería si el rey enemigo escapa por las casillas negras?''. O ``Usted afirma que cambiar damas resuelve el problema; ¿puede demostrar que ese final es ganador?''. Incluso si el alumno no propuso nada, el tutor introduce la antítesis en forma de pistas que retan: ``Parece difícil atacar por la columna, ¿existirá otra debilidad en la posición que pueda explotar?''. Esta fase es esencialmente de exploración guiada: el alumno prueba mentalmente sus hipótesis contra las preguntas del tutor.
    
    \item \textbf{Fase 3: Construcción de la solución (Síntesis):} A través del intercambio dialéctico, el alumno (con guía) va modificando su pensamiento hasta llegar a una síntesis, es decir, a una comprensión o solución que supera las limitaciones iniciales integrando las lecciones aprendidas durante la indagación. En términos prácticos, esta es la fase en que se llega a la solución del problema o al concepto buscado. Puede ocurrir de forma gradual, con el tutor reconduciendo al alumno paso a paso: ``Exacto, el rey escaparía por negras. Entonces, ¿cómo impedimos eso? --- Quizás necesito un alfil en esas diagonales. --- Bien, ¿y dispone de un alfil? --- Sí, pero está bloqueado\ldots''. Finalmente, el estudiante puede concluir, por ejemplo: ``Entonces primero tengo que redirigir mi alfil a esa diagonal y luego la torre da mate'' -- alcanzando así la síntesis, combinando su idea original con el nuevo elemento descubierto (el rol del alfil). En otros casos la síntesis surge como un momento de comprensión súbita donde el alumno, tras relacionar los elementos, anuncia la respuesta correcta. Lo importante es que esta solución se construyó con esfuerzo del alumno, no fue entregada directamente. El tutor en esta fase confirma y refuerza: ``¡Correcto! La clave era incluir al alfil para cerrar la vía de escape, y así su plan de torre funciona.''. También se clarifican detalles o se formaliza el concepto en lenguaje técnico: ``Ha aplicado un motivo de mate conocido: la batería torre-alfil que confina al rey'', dando nombre a la síntesis lograda.

    \item \textbf{Fase 4: Reflexión y abstracción:} Una vez resuelto el desafío inmediato, el ciclo continúa con una etapa de metacognición. Aquí se invita al alumno a reflexionar sobre lo ocurrido: ``¿Por qué su primera idea no era suficiente? ¿Qué aprendimos de este problema?''. Se anima a resumir la solución en generalizaciones: ``Siempre que el rey tenga casillas de escape, es necesario bloquearlas antes de dar mate'' -- extrayendo un principio. O ``Este problema nos enseñó que un plan de ataque puede necesitar la colaboración de varias piezas, no basta una sola''. Esta fase consolida la síntesis elevándola a conocimiento transferible. Además, permite al alumno tomar conciencia de su propio proceso: ``Al principio me apresuré a atacar sin ver la defensa del rival; ahora comprendo que debo verificar las respuestas del oponente, eso es importante.''. El facilitador complementa, vincula con teoría formal (si aplica), o corrige alguna conceptualización si fuera necesario. Esta es la instancia ideal para registrar en una bitácora o portafolio los hallazgos, ya que expresar por escrito (o verbalmente en clase) la lección aprendida fija mejor el aprendizaje.

    \item \textbf{Fase 5: Adaptación y próximo desafío:} La última fase cierra el ciclo actual y prepara el siguiente. Consiste en adaptar la instrucción según el desempeño observado en el ciclo que concluye, y proponer el siguiente reto acorde a ello. Si el alumno logró la síntesis con relativa facilidad, el facilitador puede decidir incrementar la dificultad o complejidad del próximo EAPBD, introduciendo quizá un tema nuevo o un problema con un patrón combinado. Si, por el contrario, el alumno necesitó mucha ayuda o aún quedó inseguro, se reforzará el mismo tema con otro ejercicio similar o se revisarán componentes más básicos antes de avanzar.
\end{itemize}

Una vez hecho el ajuste, se presenta el siguiente escenario problemático, iniciando de nuevo la fase 1 del ciclo con la nueva tesis a enfrentar. Así, el proceso es cíclico y acumulativo: cada vuelta del ciclo dialéctico-adaptativo construye sobre las anteriores, con creciente nivel de sofisticación. A lo largo de varias iteraciones, el estudiante va adquiriendo conocimientos (patrones, principios) mediante síntesis sucesivas y, paralelamente, va mejorando en las estrategias de aprendizaje (aprende a dialogar mejor, a anticipar antítesis él mismo, a reflexionar con más profundidad).

Este ciclo dialéctico-adaptativo se inspira en modelos clásicos de aprendizaje experiencial, pero integrando la esencia de la dialéctica maestro-alumno y la adaptación individualizada. En cierto modo, también recuerda al enfoque de evaluación formativa continua donde la idea de enseñar algo, observar el progreso, retroalimentar, ajustar la enseñanza, y volver a enseñar algo nuevo en un ciclo constante, solo que aquí la retroalimentación es altamente interactiva y la ``enseñanza nueva'' surge de resolver problemas. El componente dialéctico garantiza calidad en la comprensión de cada ciclo; el componente adaptativo garantiza progresión óptima a través de los ciclos.

En la práctica de aula, un maestro aplicando Odychess orquestará este ciclo con uno o varios alumnos simultáneamente. Por ejemplo, en una clase grupal, podría presentarse un EAPBD general, dejar a todos pensar (fase 1 individual), luego moderar una discusión donde surgen varias propuestas de alumnos (tesis múltiples) y confrontarlas entre ellos y con el profesor (fase 2 grupal), construir colectivamente la solución (fase 3, donde quizás diferentes estudiantes aportan elementos de la síntesis), luego pedir a algún alumno que resuma la idea (fase 4, grupal) y finalmente, según cómo percibió al grupo, proponer otra posición más compleja o aclarar sub-conceptos (adaptación fase 5). En un entorno uno a uno (por ejemplo, un estudiante practicando con la plataforma IA), el mismo ciclo ocurre, pero la interacción es estudiante vs. tutor virtual; la IA formula las preguntas dialécticas, el estudiante responde, etc., y la IA decide la siguiente tarea en base al desempeño.

El carácter cíclico e iterativo de este proceso asegura que el aprendizaje no se estanque. Incluso tras dominar un tema, el estudiante siempre se ve envuelto en un nuevo desafío que vuelve a poner a prueba y ampliar sus habilidades, manteniendo vivo el interés y la mejora continua. Además, el esquema dialéctico-adaptativo es inherentemente inclusivo: cada alumno recorre los ciclos a su propio ritmo, recibiendo en cada vuelta los estímulos y apoyos que necesita. Esto minimiza tanto la frustración del que progresa más lentamente (pues se ajusta a él) como el aburrimiento del que avanza más rápido (pues se le reta más). En síntesis, el ciclo dialéctico-adaptativo es la columna vertebral del método Odychess, combinando en la práctica los principios teóricos (dialéctica, constructivismo, andamiaje, ABP, adaptabilidad) en un proceso dinámico de aprender ajedrez.

\subsection{Adaptación del modelo de lenguaje Llama 3.3 mediante adaptación específica para tutoría personalizada de ajedrez}

Para maximizar la eficacia del tutor IA dentro del enfoque Odychess, se realizó un proceso de adaptación específica o \textit{fine-tuning} del modelo de lenguaje base Llama 3.3. La adaptación específica es una técnica de transferencia de aprendizaje en el procesamiento del lenguaje natural donde un modelo pre-entrenado a gran escala (entrenado en vastos corpus de texto general para adquirir comprensión lingüística y conocimiento del mundo) se entrena adicionalmente sobre un conjunto de datos más específico del dominio o tarea objetivo (Ruder, 2021). A diferencia del pre-entrenamiento, que requiere enormes recursos computacionales, la adaptación específica ajusta los parámetros (pesos) del modelo preexistente para especializarlo.

En nuestra implementación, se seleccionó como modelo base Llama 3.3, específicamente la versión con 70 mil millones (70B) de parámetros, debido a su demostrada capacidad de razonamiento y seguimiento de instrucciones complejas, así como su arquitectura potencialmente abierta que facilita la personalización (Touvron et al., 2023). El objetivo de la adaptación específica fue doble: primero, dotar al modelo de un conocimiento ajedrecístico preciso y fiable, asegurando la comprensión de reglas, notación (PGN, FEN), conceptos tácticos y estratégicos, y la correcta evaluación de posiciones básicas, minimizando las generaciones incorrectas o inconsistentes; segundo, inculcar el estilo pedagógico Odychess, entrenándolo para priorizar el método socrático, adaptar la dificultad y mantener un tono adecuado.

El proceso de adaptación específica se llevó a cabo utilizando técnicas de Parameter-Efficient Fine-Tuning (PEFT), las cuales son particularmente adecuadas para adaptar modelos de gran tamaño como Llama 3.3 en entornos con recursos computacionales limitados (Houlsby et al., 2019). Concretamente, se empleó la metodología Low-Rank Adaptation (LoRA) (Hu et al., 2021). La elección de LoRA fue estratégica: reduce drásticamente los requisitos de memoria al entrenar únicamente un pequeño subconjunto de parámetros adaptativos (introduciendo matrices de bajo rango en las capas del transformer y congelando la mayoría de los pesos pre-entrenados), lo cual es crucial para trabajar con GPUs de memoria limitada (típicamente 12-16GB) disponibles en plataformas accesibles como Google Colab. Además, LoRA acelera significativamente el tiempo de entrenamiento comparado con el ajuste completo del modelo (\textit{full fine-tuning}) y genera adaptadores de tamaño reducido (del orden de megabytes en lugar de gigabytes), facilitando su almacenamiento y despliegue. Fundamentalmente, esta técnica ayuda a preservar el vasto conocimiento general del modelo base (evitando el olvido catastrófico) mientras se especializa eficazmente en la tarea deseada, logrando un rendimiento comparable al ajuste completo para dominios específicos como nuestra metodología de ajedrez.

La implementación práctica se realizó en un entorno basado en la nube, específicamente Google Colab, aprovechando sus instancias equipadas con GPU (se utilizaron aceleradores como T4 o P100) y configurando entornos de ejecución con alta disponibilidad de RAM. Para manejar eficientemente la memoria, se aplicaron técnicas de cuantización, utilizando librerías como \textit{bitsandbytes}, que permiten cargar y operar el modelo con menor precisión numérica (por ejemplo, con representaciones de 4 bits o 8 bits) sin una pérdida significativa de rendimiento para la tarea de inferencia y entrenamiento adaptativo. Asimismo, se empleó la acumulación de gradientes para simular tamaños de lote (\textit{batch}) mayores a los que permitiría la memoria de la GPU, estabilizando así el proceso de entrenamiento. Todo el proceso se orquestó utilizando el ecosistema de herramientas de Hugging Face, incluyendo las librerías \textit{transformers} para la gestión del modelo, \textit{peft} para la implementación de LoRA, y \textit{datasets} para el manejo del corpus de entrenamiento (Wolf et al., 2020).

La generación del corpus de entrenamiento fue un paso crítico y multifacético, diseñado para capturar la esencia del enfoque Odychess (figura 2). Aunque las técnicas PEFT pueden funcionar con conjuntos de datos relativamente pequeños (del orden de cientos de ejemplos de alta calidad), para asegurar la profundidad pedagógica y la robustez ajedrecística deseadas, se compiló un corpus extenso de aproximadamente 50.000 ejemplos. Este corpus incluyó:

\begin{enumerate}
    \item \textbf{Diálogos Socráticos Curados:} Generados manualmente por Maestros de ajedrez y pedagogos, siguiendo los principios Odychess para la resolución guiada de EAPBDs, análisis de posiciones y explicación de conceptos, etiquetados por dificultad y competencias.
    
    \item \textbf{Partidas Comentadas con Enfoque Dialéctico:} Partidas en PGN enriquecidas con comentarios simulando interacciones Odychess (preguntas reflexivas, exploración de alternativas), verificadas tácticamente con motores como Stockfish pero con énfasis discursivo pedagógico.
    
    \item \textbf{Generación Sintética Supervisada:} Datos adicionales generados usando un modelo fundacional avanzado (como GPT-4o o Claude 3 Opus) instruido para actuar como ``Tutor Odychess'', rigurosamente filtrados y revisados por expertos humanos.
\end{enumerate}

\begin{figure}[ht]
    \centering
    \includegraphics[width=\textwidth]{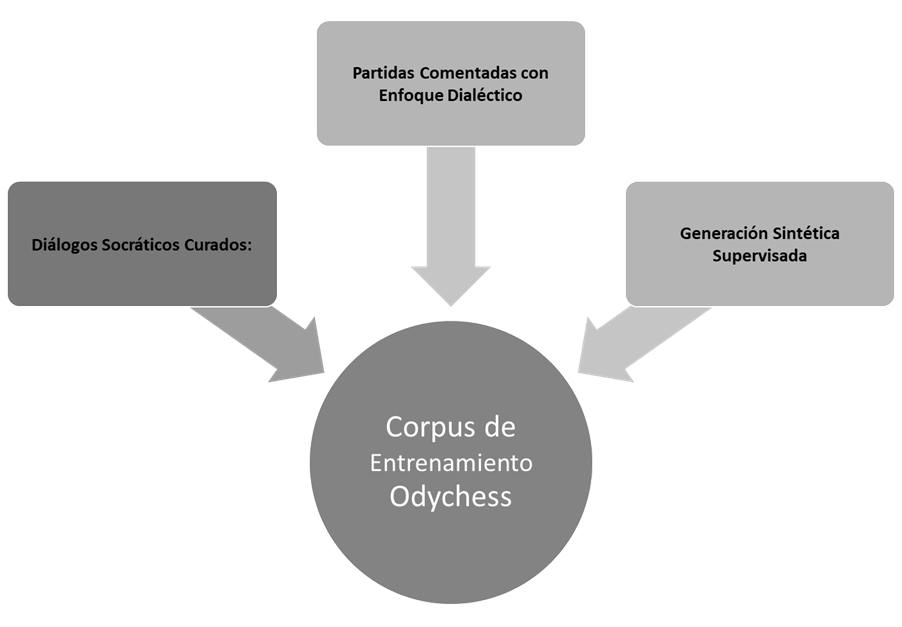}
    \caption{Componentes del corpus de entrenamiento para el Odychess-Tutor}
    \label{fig:corpus}
\end{figure}

Todos estos datos se unificaron en un formato consistente (instrucción-respuesta o diálogo continuo) adecuado para el Supervised Fine-Tuning (SFT). El entrenamiento se realizó durante varias épocas (\textit{epochs}), monitorizando la métrica de pérdida (\textit{loss}) en un conjunto de validación y evaluando periódicamente el rendimiento del modelo tanto con métricas automáticas (\textit{perplexity}, BLEU) como mediante evaluación humana experta (Raffel et al., 2020). Se ajustaron hiperparámetros clave como la tasa de aprendizaje, el tamaño efectivo del lote (mediante acumulación de gradientes) y los parámetros específicos de LoRA (r, \textit{alpha}) hasta lograr un desempeño satisfactorio. La evaluación final incluyó pruebas específicas para medir la capacidad del modelo para aplicar la metodología Odychess en la resolución de problemas ajedrecísticos novedosos, no presentes en el conjunto de entrenamiento.

\subsection{Objetivos generales y competencias a desarrollar}

El enfoque Odychess persigue un conjunto de objetivos educativos generales que orientan su diseño curricular y didáctico. Estos objetivos se traducen en competencias específicas que el estudiante deberá ir desarrollando a lo largo del programa. A continuación, se destacan los objetivos y competencias más relevantes:

\begin{itemize}
    \item Desarrollar el pensamiento estratégico y la planificación a largo plazo.
    
    \item Potenciar la resolución de problemas y la creatividad táctica.
    
    \item Fomentar el reconocimiento de patrones y la memoria ajedrecística estructurada.
    
    \item Cultivar la reflexión metacognitiva y la autocrítica.
    
    \item Fortalecer la toma de decisiones bajo incertidumbre y la gestión emocional.
    
    \item Incorporar el uso ético y eficaz de la tecnología como herramienta de aprendizaje:
\end{itemize}

Estos objetivos y competencias se operacionalizan en la metodología Odychess mediante actividades y contenidos concretos.

\subsection{Escenarios de Aprendizaje Basado en Problemas Descontextualizados (EAPBD) -- Componente Didáctico Central}

El núcleo didáctico de Odychess lo constituyen los Escenarios de Aprendizaje Basado en Problemas Descontextualizados (EAPBD). Este concepto hace referencia a situaciones de aprendizaje diseñadas en forma de problemas o desafíos ajedrecísticos específicos, presentados deliberadamente fuera del contexto de una partida convencional completa, con el fin de focalizar la atención del alumno en ciertos aspectos cognitivos o conceptuales. A diferencia de un problema tradicional de ajedrez (por ejemplo, ``mate en 2'' extraído de un final de partida conocido), los EAPBD de Odychess pueden ser posiciones, ejercicios o incluso minijuegos ajedrecísticos que no provienen necesariamente de partidas reales, o cuyo entorno narrativo ha sido abstraído, para centrar el aprendizaje en la resolución pura del problema.

Los análisis estadísticos (tabla 1) revelaron diferencias significativas a favor del Grupo Experimental (Odychess).

\begin{table}[ht]
    \centering
    \caption{Resultados Cuantitativos Clave del Estudio (Comparación Grupo Experimental vs. Grupo de Control)}
    \label{tab:resultados}
    \begin{tabular}{lll}
    \toprule
    \textbf{Variable Evaluada} & \textbf{Estadístico y p-valor} & \textbf{Tamaño del Efecto} \\
    \midrule
    PCA (Conocimiento Ajedrecístico Total) & t(58)=4,82, p $<$ 0,001 & d = 0,79 \\
    ECE (Comprensión Estratégica) & F(1, 57)=15,6, p $<$ 0,001 & $\eta^2$ parcial = 0,22 \\
    EHMA (Habilidades Metacognitivas) & t(58)=3,91, p $<$ 0,001 & d = 0,64 \\
    CMC (Motivación y Compromiso) & t(58)=5,25, p $<$ 0,001 & d = 0,85 \\
    \bottomrule
    \end{tabular}
\end{table}

El GE mostró ganancias significativamente mayores que el GC en la Prueba de Conocimiento Ajedrecístico total (t(58)=4,82, p $<$ 0,001, d = 0,79), y específicamente en la subescala de Comprensión Estratégica evaluada por la ECE (F(1, 57)=15,6, p $<$ 0,001, $\eta^2$ parcial = 0,22, controlando el \textit{pre-test}). También se encontraron diferencias significativas en la Escala de Habilidades Metacognitivas (t(58)=3,91, p $<$ 0,001, d = 0,64), indicando mayor desarrollo de la reflexión sobre el propio aprendizaje en el grupo Odychess. El análisis cualitativo de las interacciones con el Odychess-Tutor y los portafolios del GE evidenció un uso frecuente de razonamiento dialéctico (consideración de alternativas, justificación de jugadas, respuesta a contraargumentos) y una mayor profundidad en el análisis de partidas comparado con los trabajos del GC. Finalmente, el GE reportó niveles significativamente más altos de motivación intrínseca y compromiso percibido en el CMC (t(58)=5,25, p $<$ 0,001, d = 0,85).

\begin{itemize}
    \item \textbf{Consideraciones Éticas:} Se garantizó la confidencialidad y anonimato de los participantes de acuerdo con los principios éticos de investigación educativa (BERA, 2018). Los datos se almacenaron de forma segura siguiendo protocolos de protección de datos y se utilizaron exclusivamente para fines de investigación. Se informó a los participantes que podían retirarse del estudio en cualquier momento sin penalización. Al finalizar el estudio, se ofreció al Grupo de Control acceso a materiales introductorios del Enfoque Odychess para garantizar la equidad en las oportunidades de aprendizaje.
\end{itemize}

Estos resultados sugieren que la aplicación del Enfoque Odychess, mediado por un docente facilitador y el tutor IA especializado (Odychess-Tutor), fue más eficaz que la enseñanza tradicional para mejorar no solo el conocimiento y la habilidad ajedrecística (especialmente la comprensión estratégica), sino también para desarrollar competencias metacognitivas y fomentar una mayor motivación y compromiso. Es importante señalar, sin embargo, las limitaciones del estudio, como la falta de asignación aleatoria y el tamaño muestral moderado, que sugieren la necesidad de investigaciones adicionales para confirmar la generalización de estos hallazgos.

En suma, la evaluación en Odychess es un proceso rico, dinámico y centrado en el aprendizaje: informa al estudiante de su camino, informa al docente para guiar mejor, e informa al currículo para evolucionar. Al ser parte natural de cada sesión (y no un evento separado y a menudo percibido como punitivo), los estudiantes la asumen con naturalidad e incluso con interés, pues ven en ella una ayuda para ser mejores jugadores y aprendices, no un juicio estático sobre su persona. Esto contribuye a un entorno donde el foco está en el progreso individual y colectivo, y donde cada logro ---sea una nueva estrategia dominada o una actitud mejorada ante la derrota--- es reconocido como valioso.

\section{CONCLUSIONES}

El Enfoque Odychess emerge como una propuesta pedagógica original y sólidamente fundamentada, que integra la vanguardia de la teoría educativa con las significativas posibilidades que ofrece la inteligencia artificial generativa para transformar la enseñanza del ajedrez. A lo largo de este artículo, se ha detallado su diseño metodológico y didáctico, y se han presentado los resultados de una investigación aplicada que valida empíricamente su efectividad. La convergencia de principios constructivistas, la dialéctica socrática, el aprendizaje basado en problemas y patrones, y la personalización adaptativa facilitada por un LLM especializado (Odychess-Tutor) ha demostrado ser altamente beneficiosa.

La originalidad de Odychess reside en su síntesis integradora de elementos clave: se apoya en fundamentos pedagógicos bien establecidos, como el constructivismo y el poder formativo del diálogo crítico, articulándolos en un contexto contemporáneo donde las IAs avanzadas actúan como agentes educativos coadyuvantes. La investigación aplicada confirma que esta articulación, en la que el estudiante dispone de un tutor virtual inteligente bajo la guía de un docente experto, proporciona una experiencia de aprendizaje personalizada, altamente interactiva y rica en retroalimentación, superando en aspectos cruciales a los métodos tradicionales. Esto revela un potencial considerable para democratizar y elevar el nivel de la formación ajedrecística, al acercar una enseñanza de calidad casi individualizada a un mayor número de estudiantes.

La aplicabilidad del enfoque, ahora respaldada por evidencia empírica, se proyecta amplia. Sus principios son transferibles a otros dominios del conocimiento que valoran el aprendizaje por descubrimiento guiado y la personalización. En el ámbito específico del ajedrez, Odychess ha demostrado ser viable y efectivo en programas escolares para el desarrollo de habilidades transversales, y se vislumbra su utilidad en academias de alto rendimiento o plataformas en línea. La viabilidad técnica de la adaptación específica de modelos como Llama 3.3, mediante técnicas eficientes como LoRA en plataformas accesibles, quedó demostrada durante la preparación del estudio, y su impacto positivo en el aprendizaje resultó significativo.

La naturaleza dialéctica y adaptativa de Odychess sugiere que futuras implementaciones y ciclos de investigación-acción continuarán refinando el método, los EAPBDs y el propio tutor IA, optimizando progresivamente su robustez y efectividad. Odychess se presenta, por tanto, no como un método estático, sino como un marco flexible y validado que invita a una adopción y adaptación informada.

En conclusión, el enfoque Odychess se distingue por su concepción integradora y sinérgica de pedagogía y tecnología, respaldada por la evidencia empírica; su efectivo enfoque en el desarrollo del pensamiento estratégico y metacognitivo, más allá de la mera instrucción ajedrecística; su probada adaptabilidad a distintos niveles dentro del rango estudiado, gracias al andamiaje progresivo y la inteligencia artificial; y su carácter transformador evidenciado en los resultados del estudio, que apoyan la visión de ``un tutor para cada alumno'' como una vía prometedora para redefinir la educación.

La implementación exitosa de este enfoque en el estudio sugiere la posibilidad de formar una nueva generación de aprendices de ajedrez que no solo mejoren su juego, sino que también desarrollen un pensamiento más profundo y reflexivo.

El impacto potencial de Odychess trasciende el tablero de 64 casillas, contribuyendo al ideal de una educación apoyada en la IA que potencie, en lugar de reemplazar, la creatividad y el intelecto humanos. Este artículo, enriquecido con la metodología y los resultados de la investigación, constituye una base sólida para el desarrollo de manuales de aplicación y guías docentes, invitando a futuras investigaciones y aplicaciones que continúen explorando y refinando este prometedor enfoque.

\section*{REFERENCIAS BIBLIOGRÁFICAS}

Barrows, H. S. (1986). A taxonomy of problem-based learning methods.
\textit{Medical Education}, 20(6), 481-486.\\
\\
Braun, V., \& Clarke, V. (2006). Using thematic analysis in psychology.
\textit{Qualitative Research in Psychology}, 3(2), 77-101.\\
\\
Bueno Pérez, L. A. (2000). \textit{Modelo para la enseñanza-aprendizaje del
ajedrez en la universidad}. Tesis de Doctorado (Doctorado en Ciencias
Pedagógicas). Santiago de Cuba, Universidad de Oriente. Centro de
Estudios de la Educación Superior ``Manuel F. Gran''.\\
\\
Bueno, L. (2015). \textit{Ajedrez juego ciencia y conciencia}. Editorial
Academia. La Habana, Cuba.
\url{https://www.libreriavirtual.cu/libreria/ajedrez-juego-ciencia-y-conciencia}\\
\\
Chase, W. G., \& Simon, H. A. (1973). Perception in chess. \textit{Cognitive
Psychology}, 4(1), 55-81.\\
\\
Coll, C., Martín, E., Mauri, T., Miras, M., Onrubia, J., Solé, I., \&
Zabala, A. (2006). \textit{El constructivismo en el aula}. Graó.\\
\\
De Groot, A. D. (1965). \textit{Thought and choice in chess}. Mouton.\\
\\
Gobet, F., \& Simon, H. A. (1996). Templates in chess memory: A mechanism
for recalling several boards. \textit{Cognitive Psychology}, 31(1), 1-40.\\
\\
Grattafiori, A., Dubey, A., Jauhri, A., Pandey, A., Kadian, A.,
Al-Dahle, A., \ldots \& Vasic, P. (2024). The llama 3 herd of models.
\textit{arXiv preprint arXiv:2407.21783}.\\
\\
Ramírez-Guerra, D. M., Bueno-Pérez, L. A., \& Gordo-Gómez, Y. M.
(2016). La capacitación en el proceso de masificación del ajedrez en
las comunidades urbanas. \textit{Ciencia y Deporte}, 1(2), 56-71.\\
\\
Reyes-Joa, H. M., Ramirez-Guerra, D. M., \& Bueno-Pérez, L. A. (2020).
El perfeccionamiento del proceso de enseñanza-aprendizaje del Ajedrez
de la categoría 10-11 años. \textit{Ciencia y Deporte}, 5(1), 19-31.\\
\\
Savery, J. R. (2006). Overview of problem-based learning: Definitions
and distinctions. \textit{Interdisciplinary Journal of Problem-Based Learning},
1(1), 9-20.\\
\\
Serper, G. (2015). Patterns, patterns everywhere! Chess.com.
\url{https://www.chess.com/article/view/chess-patterns-patterns-everywhere}\\
\\
Wolf, T., Debut, L., Sanh, V., Chaumond, J., Delangue, C., Moi, A.,
Cistac, P., Rault, T., Louf, R., Funtowicz, M., et al. (2020).
Transformers: State-of-the-art natural language processing. \textit{Proceedings
of the 2020 Conference on Empirical Methods in Natural Language
Processing: System Demonstrations}, 38-45.\\
\\
USFA (2024). Inteligencia Artificial Generativa en la Educación:
Innovación con Conciencia.
\url{https://www.usfa.edu.bo/blog/noticias/post/inteligencia-artificial-generativa-en-la-educacion-innovacion-con-conciencia}\\

\end{document}